\documentstyle[aps,prb,epsfig]{revtex}
\begin{document}
\draft
\title
{Physical picture of the gapped excitation spectrum of the 
one-dimensional Hubbard model }
\author{Daniel Braak$^{1,2}$ and Natan Andrei$^{1}$}
\address{$^1$ Department of Physics and Astronomy, Rutgers University,
Piscataway, NJ 08855}
\address{$^2$ NEC Research Institute, 4 Independence Way, Princeton,
NJ 08540}
\date{\today}
\maketitle
\begin{abstract}
 A simple picture for the  spectrum of the one-dimensional Hubbard
  model
is presented using a classification of the 
eigenstates based on an  intuitive bound-state Bethe-Ansatz approach. 
This approach allows us to {\it prove} a "string hypothesis" for complex
 momenta and derive an exact formulation of the Bethe-Ansatz equations
including all states. Among other things we
  show that all gapped eigenstates  have the Bethe Ansatz form,
 contrary to assertions in the literature \cite{ess}. The 
 simplest  excitations in the upper Hubbard band are computed: we find  
an unusual
 dispersion close to half-filling.  
\end{abstract}
\def\p{\phi}
\def\P{\Phi}
\def\e{\eta}
\def\eps{\epsilon}
\def\vep{\varepsilon}
\def\ps{\psi}
\def\a{\alpha}
\def\ab{{\tilde{a}}}
\def\b{\beta}
\def\y{{\tilde{y}}}
\def\k{\kappa}
\def\psd{\ps^{\dagger}}
\def\psdt{{\tilde{\ps}}^{\dagger}}
\def\pst{\tilde{\ps}}
\def\di{d^{\dagger}}
\def\d{\delta}
\def\g{\gamma}
\def\be{\begin{equation}}
\def\ee{\end{equation}}
\def\bdm{\begin{displaymath}}
\def\edm{\end{displaymath}}
\def\s{\sigma}
\def\bea{\begin{eqnarray}}
\def\eea{\end{eqnarray}}
\def\bear{\begin{array}}
\def\eear{\end{array}}
\def\nn{\nonumber}
\def\x{\chi}
\def\ra{\rightarrow}
\def\r{\rho}
\def\s{\sigma}
\def\vs{\vec{\s}}
\def\S{{\hat{S}}}
\def\Sr{\tilde{S}}
\def\z{\zeta}
\def\Z{\bar{Z}}
\def\Zr{\tilde{Z}}
\def\ta{\tilde{A}}
\def\t{\tau}
\def\th{\theta}
\def\Th{\Theta}
\def\l{\lambda}
\def\L{\Lambda}
\def\K{{\cal K}}
\def\Li{{\cal L}}
\def\E{{\cal E}}
\def\ki{\frac{k_i}{\L}}
\def\kj{\frac{k_j}{\L}}
\def\ua{\uparrow}
\def\da{\downarrow}
\def\rd{\textrm{d}}
\def\id{{\mathbf{1}}}
\def\ph{\phantom}
\def\O{\Omega}
\def\o{\omega}
\def\H{\tilde{H}}
\def\N{{\hat{N}}}
\def\arsinh{{\textrm{arsinh}}}
\def\cotanh{{\textrm{cotanh}}}
\def\arcsin{{\textrm{arcsin}}}
\def\dim{{\textrm{dim}}}
\def\sech{{\textrm{sech}}}
\def\vu{\frac{4}{u}}
\def\uv{\frac{u}{4}}
\def\uz{\frac{u}{2}}
\def\cz{\frac{c}{2}}
\def\dx{\Delta\x}
\def\ecl{e^{-\frac{|c|}{2}L}}
\def\eclp{e^{\frac{|c|}{2}L}}
\def\eul{e^{-\frac{|u|}{4}L}}
\def\eulp{e^{\frac{|u|}{4}L}}
\def\eklz{e^{-\xi_l L}}
\def\ekl{e^{-\xi L}}
\def\eklp{e^{\xi L}}
\def\ekld{e^{-2\xi L}}
\def\ln{{\textrm{ln}}}
\def\G{{\hat{G}}}
\def\ran{\rangle}

\section{Introduction }

 In a recent article \cite{bran} we have introduced a new approach to
the derivation of the Bethe Ansatz equations of models 
with bound states. By bound states we  mean that the wave functions
contain complex momenta $k_j$ which need to be properly handled in the infinite
volume limit. Our approach avoids postulating a "string hypothesis" for the
 momenta and allows a systematic and simple procedure to derive all 
eigenstates.

 In our approach  the Bethe Ansatz is formulated for  bound states (and bound states of
bound states). These composites are formed on the infinite line
and are incorporated systematically
 as building blocks of
 the wave functions
 defined  on a finite ring. The composites
correspond to  {\it poles} of the respective S-matrices thus guaranteeing their stability
under scattering with other particles. 
This necessitates the introduction of  appropriate
 boundary conditions - Composite Boundary Conditions (CBC) -  which respect
 the 
construction and allow the formation of {\it exact strings}.
 We conjecture that the CBC  allow a  symmetry - presumably Yangian - 
to be manifest already 
at finite
 volume, a symmetry that is  violated by the conventional 
Periodic Boundary Conditions, hence appearing in the latter case only
in the infinite volume limit. We shall refer to this approach as the Bethe Ansatz for Composites (BAC).

The approach is general and applies to any model where complex momenta appear:
 the Hubbard model, the t-J model, the Anderson model and the multichannel Kondo
model among many others.
We shall discuss here the Hubbard model in detail and leave the treatment of 
the other models for later.

The article is organized as follows:

In section II we present the model and 
 construct the class of bound  state
solutions to the Schr\"odinger equation of the
 model and show how to incorporate them as
new ansatz functions besides the usual plane waves into the
BA. We introduce appropriate boundary conditions 
(the  CBC) to define the
model on a finite configuration space, and deduce the BAC equations.
A particular class of solutions of the BAC equations is found to underlie
the $\eta$-pairing and the charge $SU(2)$ symmetry group.
Next, we clarify the connection with Takahashi's
string hypothesis \cite{taka}, which, we  argue, leads to inconsistencies when
finite volume corrections are taken into
account. 

In section III we study the bound state excitation spectrum both in the
repulsive and  attractive models. In the repulsive case, holding
the number of electrons fixed, we find a three parameter excitation,
consisting of two gapless {\it holons} (spinless, carrying charge -$e$)
and a bound state residing in the upper Hubbard band (spinless, carrying 
charge 2$e$). The latter is in fact an anti-bound state, corresponding
 to the formation of a bound pair of electrons
with positive binding energy  dressed by its interaction with the sea
electrons. In the attractive case a dual picture emerges:
it corresponds to breaking
one of the pairs forming the ground state.  The resulting excitation
consists of two gapless (renormalized)
 electrons each carrying charge $e$ and spin $1/2$
and a gapped spinless excitation with charge -$2e$.
The gapped excitation is an 
independent mode only away from half-filling and we give numerical
results for its dispersion.

In section IV we summarize our results.
Appendix A contains a detailed demonstration of the
stability of the two-particle bound state in the presence of
other particles with composite boundary conditions.

\section{Derivation of the BAC equations}
The one-dimensional Hubbard hamiltonian is given by
\be
H=\sum_{i=-\infty}^{\infty}-t(\psd_{\s,i+1}\ps_{\s,i} +
h.c.)+Un_{\ua,i}n_{\da,i}.
\label{sham}
\ee
The hamiltonian was diagonalized by Lieb and Wu \cite{liebwu}
in the sector with $N$ particles  on a finite ring of length $L$, with 
 periodic boundary
conditions (PBC) imposed on the wavefunction of the $N$ electrons,
\be
F(x_1,\ldots, x_i,\ldots,x_N)=F(x_1,\ldots, x_i+L,\ldots,x_N)\ \ \ \
\forall\  i.
\label{pbc}
\ee

The resulting eigenfunction is parameterized, for total spin
 $S=\frac{1}{2}(N-2M)$, by
$N$ momenta $k_j$ 
and $M$ spin rapidities $\l_\g$, satisfying
 the BA equations \cite{liebwu,and},
\be
e^{ik_jL}=\prod_{\d=1}^M\frac{\l_\d-\sin k_j-i\frac{u}{4}}
{\l_\d-\sin k_j+i\frac{u}{4}}
\label{lw1}
\ee
\be
\prod_{\d\neq\g}^{M}\frac{\l_\g-\l_\d-i\frac{u}{2}}
{\l_\g-\l_\d+i\frac{u}{2}}
=\prod_{j=1}^{N}\frac{\l_\g-\sin k_j-i\frac{u}{4}}
{\l_\g-\sin k_j+i\frac{u}{4}}
\label{lw2}
\ee
where $u=U/t$. 

 The energy and momentum of
the state are then given by
\[
E=-2t\sum_j \cos k_j \ \ \ \ P=\sum_j k_j. 
\]
The eigenstates of the hamiltonian correspond to the various solutions of 
eqns (\ref{lw1},\ref{lw2}). Real as well as complex solutions need to be 
considered to obtain a complete spectrum of states.

We shall argue, however, that complex spin rapidities $\{\l_\g \}$ and complex 
charge momenta $\{k_j\}$ have a different character
although both of them were treated in much the same way in the
literature. 
The former type is associated with the spin degrees of freedom
and describes kink/anti-kink bound states. As such they are
a many-body phenomenon and make their appearance through
complex conjugate solutions of the BA equations for
spin rapidities. The existence of these solutions is
postulated in the  $\l$ {\it string-hypothesis} about
the form of the BA solutions for large number of
particles $N$. Although very plausible, this hypothesis remains
unproven up to now \cite{abd}.

The latter type - on which we concentrate in this article - 
is associated with the charge degrees of freedom.
In analogy to the spin sector, they were assumed
to correspond to complex conjugate solutions of the 
Lieb - Wu equations for charge rapidities. We will show that they
have to be treated differently because they are {\it not} due to 
a many-body effect but can be identified with
the {\it elementary} bound states of the hamiltonian, 
 present already in the few-particle sector.
They have to be incorporated into the Bethe ansatz
ab initio and lead to a set of BA equations, which
we call the Bethe ansatz for composites (BAC).

 The charge complex momenta were also conjectured to form strings, the so
called  $k - \L$ {\it strings},  and Bethe-Ansatz
 equations were derived based on this
hypothesis \cite{taka}.
 In the thermodynamic
limit the BAC equations turn out to be equivalent with
those derived using the string-hypothesis for the charge
rapidities. This finding does not constitute, however, a proof of 
the string-hypothesis for a finite system. On the contrary, we shall
argue 
in section II D. that
the
string-hypothesis for charge bound states leads to
an {\it over-constrained} set of equations, which  has
in general no solutions for sufficiently large but finite system size
$L$. 

\bigskip

We begin by discussing the two string hypotheses separately.
To introduce "$\l$ - strings"
one assumes  complex-conjugate pairs
of rapidities
 $\l^\pm=\l_0\pm i\chi$, and introducing them into eq
(\ref{lw2}) one concludes,
\be
\chi = \uv + {\cal{O}}(e^{-\kappa N})
\ee
with $\kappa>0$. More generally, $m$ spin rapidities
$\l_j$ are grouped together to form a
 $\l$-string of length $m$:
\be
\l_j=\l_0 + i (m+1-2j)\uv  \ \ \ \ \ j=1 \ldots m.
\label{spinstr}
\ee
The  solutions with complex $\l$ to the BA equations are driven to the
string position (\ref{spinstr}) in the limit $N\ra\infty$,
 corresponding to a many particle effect.

 The standard 
classification emerges in terms of {\it spinons} - chargeless spin-$1/2$ - objects. The ground state
is a singlet, $S=0$, described by  a solution with $M=N/2$ real 
$\lambda$ rapidities. The
simplest spin excitation corresponds to $M=N/2 -1$ rapidities, hence $S=1$.
It is found to be
a two parameter state and can be interpreted  naturally as a
symmetric combination  of {\it two} elementary excitations, each
carrying spin $1/2$, the {\it spinon} states. 
To confirm 
this picture
one needs to show that there is also a state with the  spinons
 combined antisymmetrically 
to form an excited singlet. This can be done
by introducing
a $\l$-string of length 2 so that again $M=N/2$ corresponding to 
a singlet, $S=0$. This structure of excitations is universal to $SU(2)$ 
invariant models.
It was first observed in the Gross-Neveu model \cite{anlo},
and subsequently rediscovered in several  other models: in the Kondo model
\cite{an}, the Heisenberg model \cite{fadta}, the Hubbard model \cite{woy}
or the t-J model \cite{bbo}. Also the spinon - spinon S-matrix, 
when expressed in terms of rapidities, is identical in
 these models, and was originally calculated from the Bethe Ansatz
 in [\onlinecite{anlo2}]. The fact that quantities such as
the quantum numbers of the spinons
and their scattering
phase shifts  are the same in the different models is due to the 
circumstance that in all of them the interaction has the form of a 
spin exchange. Thus  the spin sector of the various models
obeys the same BA equations up to  shifts specific
to each. As phase shifts and quantum numbers
 are deduced from simple counting arguments 
that are independent of those shifts they are identical in spite of
a wide variation in dynamics, see [\onlinecite{and}] for a 
detailed discussion.

\bigskip
 
In 
addition to these spin strings the standard approach assumes
``$k-\L$ strings'', which should describe the
charge bound states \cite{taka}. In the simplest case one assumes two complex
momenta $k^\pm$ are grouped together with a certain
spin rapidity
$\L$ from the set of the $\l_\g$:
\be
\sin k^\pm=\L\mp i\uv +{\cal{O}}(\eul),
\label{stri}
\ee
which one proceeds to insert into the Lieb - Wu equations. This procedure 
is assumed to hold for $k-\L$ strings of any length.

 We shall argue in what follows that this approach is flawed (a brief account 
was given in
[\onlinecite{bran}])  and will introduce in
its place the BAC approach, which does not rely on a 
 string-hypothesis. We shall argue later that solutions of 
type (\ref{stri}) {\it do not} exist for finite $L$.
This is due to the fact that the elementary bound states
of (\ref{sham}), to be constructed below,
cannot be defined with periodic boundary conditions in the presence of unbound
electrons. Therefore, such states -  expected to exist in the infinite volume limit -  cannot be obtained from finite volume considerations. To obtain
a consistent solution at finite volume 
we need to treat the bound states on an equal footing
with the plane wave (scattering) states in the Bethe ansatz,
rather than trying to recover them as special solutions,
``$k-\L$ strings'', of the Lieb-Wu equations, which were derived
   from  a Bethe ansatz based on plane waves.

It is amusing to note that already 
the expression (\ref{stri}) hints that it should not be treated, as is conventionally done, on par with the spin strings. 
In contrast to the latter the
$k-\L$ strings are driven to their asymptotic form in the
limit $L\ra\infty$, i.e. in the infinite volume limit
without the need for $N$ to be large as well.
This suggests that they are not due to a many-body
correlation effect as the spin strings (\ref{spinstr}) and
should therefore exist already in the
two-particle sector of the Hilbert space. 

 We proceed now to explain our approach in detail.

 \subsection{  Elementary bound states and their boundary conditions}

 The Schroedinger equation of the Hubbard model for
$N$ particles reads,
\bea 
 \sum_{i}^NF_{a_1\ldots a_N}(n_1,\ldots,n_i-1,\ldots,n_N) 
&+&F_{a_1\ldots a_N}(n_1,\ldots,n_i+1,\ldots,n_N) \label{schroe} \\
+U\sum_{i<j}^N\d_{n_in_j}&F&_{a_1\ldots a_N}(n_1,\ldots,n_N) 
 =EF_{a_1\ldots a_N}(n_1,\ldots,n_N). \nn 
\eea
The solution of (\ref{schroe}) in the two particle sector consists of:\\
(i) combination of plane waves 
$F_{a_1a_2}(n_1,n_2)= A_{a_1a_2}e^{i(k_1n_1+k_2n_2)}$ describing
unbound particles. These states  exist on the infinite line as well as
on a finite ring of length $L$,\\ 
(ii) bound state solutions which take the following form on the infinite line:
\be
F^{b}(n_1,n_2)=A^s_{a_1a_2}e^{iq(n_1+n_2)}e^{-\xi(q)|n_1-n_2|}.
\label{bst}
\ee
Here $A^s_{a_1a_2}$ denotes a spin singlet, i.e. $A^s_{a_1a_2}=-A^s_{a_2a_1}$. 
The parameters $q$ and $\xi$ are related by,
\be
\sinh\xi(q)=-\frac{u}{4\cos q}.
\label{xik}
\ee
The energy of the bound state is $E(q)=-4t\cos q\cosh\xi$ and $\xi\ge 0$.
The last condition together with (\ref{xik})
sets a range for $q$ in the interval $[-\pi,\pi]$:
\be
\bear{ll}
U>0:\ \ \ &|q|\ge\frac{\pi}{2}\\
U<0:\ \ \ &|q|\le\frac{\pi}{2}.
\eear
\ee
The signs of the energy  depends directly on
$U$: it is negative in the attractive case ($U<0$) and positive in the
repulsive case. In the former case the real momentum $q$ (which is half
the quantum mechanical (or crystal) momentum: $p=2q$) 
lies near the center of the
Brillouin zone and in the latter near the edges. Moreover there is a
gap
in the spectrum $E(q)=U\cotanh\xi(q)$: 
\be
\bear{ll}
U>0:\ \ \ &E(q)\ge |U|\\
U<0:\ \ \ &E(q)\le -|U|.
\label{UU}
\eear
\ee
The limiting values $\pm |U|$ correspond to $q\ra \pm(\pi/2)$ from above
or below according to the sign of $U$. In both cases  $\xi(q)$ tends to
infinity at these points. The state with $q=\pi/2$ (equivalent with
$q=-\pi/2$) is strictly local - with a zero width wave function
 both electrons being one on top of the other.
The wave function of the local pair is characterized by the crystal momentum $p=\pi$ and
energy $E=U$. This local bound state is the
only eigenstate of (\ref{sham}), which exists also on a finite
ring with even number of lattice sites and plays an important role
for the $SU(2)_{charge}$ symmetry, see below.

The bound-state solution (\ref{bst}) corresponds, as it should, to a pole
of the appropriate S-matrix (or a zero in its inverse). We proceed to make 
it manifest.

The  scattering matrix between two  unbound electrons,
\be
S_{ij}(k_i,k_j)=\frac{\sin k_i-\sin k_j +i\frac{u}{2}P_{ij}}{
\sin k_i-\sin k_j + i\frac{u}{2}},
\label{free}
\ee
is used to construct the conventional scattering solution,

\be
F(n_1,n_2)={\cal A}e^{i(k_1n_1+k_2n_2)}(A_{a_1a_2}\Th(n_2-n_1)
+[S_{12}A]_{a_1a_2}\Th(n_1-n_2))
\label{conven}
\ee
where  $\cal A$ denotes the antisymmetrizer and the momenta are real.

To cast expression (\ref{bst}) into this form, rewrite it
as,
\be
A^s_{a_1a_2}e^{i(q-i\xi)n_1}e^{i(q+i\xi)n_2}\Th(n_2-n_1)
+A^s_{a_1a_2}e^{i(q-i\xi)n_2}e^{i(q+i\xi)n_1}\Th(n_1-n_2),
\ee
which is of the form (\ref{conven}) if we identify:
$k_1=k^-=q-i\xi$ and $k_2=k^+=q+i\xi$, provided $[S_{12}A^s]_{a_1a_2}=0$.
Using now the bound state condition (\ref{xik}), we find, 
$\sin k^{\pm} = \sin q \cosh \xi \pm i\cos q  \sinh \xi =\sin q \cosh
\xi \mp i\frac{u}{4} \equiv \p(q)\mp i\frac{u}{4}$.
We have used the following notation:
\bea
\p(q)=\Re(\sin k^\pm)=\sin q \left(1+\frac{u^2}{16\cos^2 q}\right)^{1/2}.
\label{phiq}
\eea
Thus, inserting these values into eq.(\ref{free}) we have, 
\be
S_{12}(k^-,k^+)=\frac{1}{2}(\id+P_{12}).
\label{sings}
\ee
Observe that
$S_{12}$ vanishes indeed for the singlet ($P_{12}=-1$):
$[S_{12}A^s]_{a_1a_2}=0$, while,
$S_{21}=S_{12}^{-1}=S_{12}(k^+,k^-)$, is in 
turn  undefined \cite{yang2}: the complex momentum satisfying
the bound state condition (\ref{xik}) is placed
 on the {\it pole} -- a standard result of scattering theory.

Hence (\ref{conven}) gives correctly the  wavefunction (\ref{bst}) after
antisymmetrization as the amplitude $e^{\xi x}$ which diverges in
interval
$x>0$ is projected out:
\be
F(n_1,n_2)={\cal
A}e^{i(q-i\xi)n_1}e^{i(q+i\xi)n_2}A^s_{a_1a_2}\Th(n_2-n_1)=
A^s_{a_1a_2}e^{iq(n_1+n_2)}e^{-\xi|n_1-n_2|}.
\ee 
We conclude that
the bound state can be written in the usual Bethe ansatz form
(\ref{conven})
if we allow for singular S-matrices like (\ref{sings}).

If we seek an
analogue
of this bound state solution in the framework of PBC we find that
the momentum is not placed at the pole of the S-matrix.
Assume  two electrons on the ring with
$L$ sites. Let us consider a singlet wave function in the two-particle sector,
 $N=2$, and  $S=0$, (i.e. $M=1$). The equations (\ref{lw1},\ref{lw2}) read, 
\be
e^{ik_jL}=\frac{\l-\sin k_j-i\frac{u}{4}}
{\l-\sin k_j+i\frac{u}{4}}, \ \ j=1,2
\label{lw11}
\ee
\be
1
=\prod_{j=1}^{2}\frac{\l-\sin k_j-i\frac{u}{4}}
{\l-\sin k_j+i\frac{u}{4}}.
\label{lw21}
\ee
Now we look for solutions of (\ref{lw11},\ref{lw21})
with complex $k_j$, of the form $k_{1,2} = q \pm i \xi$. Inserting 
 into the equations we find that
  the spinon parameter $\l$ is given by
$\l=\p(q)$ and
 $\xi$ satisfies the following equation
\be
e^{-2\xi L}=\left(\frac{\cos q\sinh \xi+\uv}{\cos q\sinh
\xi-\uv}\right)^2,
\ee
which means,
\be
|\cos q|\sinh \xi-\frac{|u|}{4}\sim e^{-\xi L}
\ee
for finite $L$. It follows that the bound state momentum 
in a finite system with PBC according to the
Lieb-Wu equations deviates from the pole in the corresponding
S-matrix  by a term of order $e^{-\xi L}$. This
makes
$S_{12}$ regular and leads to a $F(n_1,n_2)$ periodic in each variable
separately, because in each sector of configuration space there is
also an amplitude with the exponentially diverging solution. These
amplitudes are forbidden only for $L\ra\infty$. The fact, that the
bound state parameter {\it does not} satisfy the pole condition
(\ref{xik}) for finite $L$  renders the
bound state unstable when scattered  on additional particles\cite{dan}.
To avoid this problem, we have to introduce boundary conditions,
which allow the state (\ref{bst}) to exist in a finite system. This is
done in the next section.

\subsection{ The Composite Boundary Conditions (CBC) and the Bethe Ansatz for
 Composites (BAC)
}
We proceed to introduce  boundary conditions which allow us to 
incorporate bound-states consistently, bypassing the unsatisfactory ``string hypothesis''.  We shall show that the Composite 
Boundary Conditions (i) define a complete Hilbert space spanned by 
$4^L$ states for finite volume $L$, (ii) lead in
in the infinite 
volume limit to the conventional quantization of
 the infinite line, (iii) respect the formation of the bound states.
We conjecture that they allow, already for 
finite volume, a symmetry expected to hold only in the infinite volume 
limit. The presence of this symmetry may underlie the {\it exact} form of our Bethe Ansatz equations.

To impose a boundary condition on particle $j$ we have to
find a path in configuration space leading from region
$x_j<x_1<x_2\ldots<x_N$ to region $x_1<x_2\ldots<x_N<x_j$. If $j$
belongs
to a bound state, the corresponding product of S-matrices is
necessarily
singular. We can avoid the singularity by taking the two members $j^-$
and $j^+$ simultaneously around the ring of circumference $L$, 
by imposing the following modified boundary condition for
bound states,
\be
F(x_1\ldots,x_{j^-},\ldots,x_{j^+},\ldots x_N)=F(x_1\ldots 
,x_{j^-}+L,\ldots,x_{j^+}+L,\ldots x_N)
\label{bsbc}
\ee
corresponding to periodicity of the center of mass coordinate. For unbound particles we retain conventional periodicity. Two new nonsingular S-matrices arise.
The S-matrix of a bound pair with an unbound particle, $S^{ub}_{i(j^-j^+)}$, and the S-matrix, $S^{bb}_{(i^-i^+)(j^-j^+)}$, between two bound pairs. These are in addition to
the usual $S^{uu}_{ij}$ given by (\ref{free}) valid for both $k_i$ and
$k_j$ real.

The factorization property of the S-matrices (\ref{free}) makes the
following construction possible:
Because of the validity of the Yang Baxter relation it is
sufficient
to consider only amplitudes in which the two members of the bound pair,
with momenta $k^{\pm} = q \pm i\xi$, are neighbors.  
Obviously the scattering matrix of the pair off an unbound particle with momentum $k$ is,
\be
S^{ub}_{1(23)}=S^{uu}_{13}(k,k^+)S^{uu}_{12}(k,k^-),
\label{frb}
\ee
and we find (see appendix A),
\be
S^{ub}_{1(23)}(k,q)=\frac{\sin k -\p(q) -i\frac{u}{4}}
{\sin k -\p(q) +i\frac{u}{4}}.
\label{sub}
\ee
$S^{ub}$ is a scalar in spin-space, and acts as a pure 
(momentum dependent) phase shift on
the
wavefunctions.
The bound state is {\it stable},  its internal  wavefunction in
 is not affected by the scattering with the unbound particle.
In other words, it couples directly
only
to the charge degrees of freedom via the phase-shift (\ref{sub}).
In an analogous way
we derive  $S^{bb}_{(12)(34)}(q_a,q_b)$ by
using,
\be
S^{bb}_{(12)(34)}=S^{ub}_{1(34)}(k_{a}^-,q_{b})
S^{ub}_{2(34)}(k_{a}^+,q_{b})
\label{bbsm1}
\ee
with the result
\be
S^{bb}_{(12)(34)}(q_a,q_b)=\frac{\p(q_a) -\p(q_b) -i\frac{u}{2}}
{\p(q_a) -\p(q_b) +i\frac{u}{2}}.
\label{bbsm}
\ee

We proceed now to derive the Bethe-Ansatz equations. As
$S^{ub}$ and $S^{bb}$ commute with each other and all the $S^{uu}$'s
 they do not enter the self consistency BA equations
(\ref{lw2}).
This fact is responsible for the (partial) decoupling of the bound states from
the free particles which are correlated via the spinon parameters $\l_\g$.
Assume $N=N^u +2N^b$ electrons, where $2N^b$ particles are in 
bound states characterized by momenta $q_l$, $l=1\ldots N^b$ and
$N^u$ particles are in plane wave states given by momenta $k_j$,
$j=1\ldots N^u$. The total spin $S$ is given by
$S=\frac{1}{2}(N^u-2M)$,
where $M$ denotes the number of spin-lowering operators in the
algebraic Bethe ansatz. 

One proceeds to construct the  $Z$-matrix which takes a particle or a 
bound pair
around the ring $L$. For a  particle
$j$ in an unbound state it takes the form,
\be
Z_j = S^{uu}_{1j}\ldots S^{uu}_{N^uj}S^{ub}_{1j}\ldots S^{ub}_{N^bj}
\ee
with eigenvalue $z_j=e^{ik_jL}$. The last $N^b$ S-matrices are
phases. Diagonalizing $Z_j$  is a standard procedure and 
 we find the BA equations,
\be
e^{ik_jL}=\prod_{\d=1}^M\frac{\l_\d-\sin k_j-i\frac{u}{4}}
{\l_\d-\sin k_j+i\frac{u}{4}}
\prod_{l=1}^{N^b}\frac{\p(q_l) -\sin k_j -i\frac{u}{4}}
{\p(q_l) -\sin k_j +i\frac{u}{4}}
\label{rek}
\ee
\be
\prod_{\d\neq\g}^{M}\frac{\l_\g-\l_\d-i\frac{u}{2}}
{\l_\g-\l_\d+i\frac{u}{2}}
=\prod_{j=1}^{N^u}\frac{\l_\g-\sin k_j-i\frac{u}{4}}
{\l_\g-\sin k_j+i\frac{u}{4}}
\label{spin}
\ee

Similarly the $Z$-matrix for a bound pair  determines  its momentum, $q_l$, 
and is given by,

\be
Z_{(ij)}=S^{ub}_{1(ij)}\ldots
S^{ub}_{N^u(ij)}S^{bb}_{(r^-_1,r^+_1)(ij)}
\ldots S^{bb}_{(r^-_{N^b},r^+_{N^b})(ij)}
\label{zbp}
\ee
with eigenvalue $z_{(ij)}=e^{2iq_lL}$.
The resulting BA equations read,
\be
e^{2iq_lL}=\prod_{j=1}^{N^u} \frac{\sin k_j -\p(q_l) -i\frac{u}{4}}
{\sin k_j -\p(q_l) +i\frac{u}{4}}
\prod_{n\neq l}^{N^b}\frac{\p(q_n) -\p(q_l) -i\frac{u}{2}}
{\p(q_n) -\p(q_l) +i\frac{u}{2}}
\label{bsk}
\ee
The equations (\ref{rek},\ref{spin},\ref{bsk}) replace the set (\ref{lw1},\ref{lw2})
in the presence of two-particle bound states.
These bound states, however, are not the only multi-particle
complexes
which are allowed by the hamiltonian. We can infer the existence of 
higher composites by looking for zeros and poles of the
S-matrix for two two-particle bound states (\ref{bbsm}). The zero of the
 bound -
state S-matrix,
(\ref{bbsm}), is at,
$\p(q_a)-\p(q_b)=i\uz$,  corresponding to complex $q$'s. 
Choosing
$\p_{1,2}^{2}=\p_0^2\pm i\uv$, we find the four momenta of this
``double'' bound state, the so-called quartet \cite{and},
\bea
\sin k^2_1=\p_1^2+i\uv \nn\\
\sin k^2_2=\p_1^2-i\uv  \nn\\
\sin k^2_3=\p_2^2+i\uv \nn\\
\sin k^2_4=\p_2^2-i\uv.
\eea
In other words,
\bea
k^2_1&=&\pi-\arcsin(\p^2_0+i\uz) \nn\\
k^2_2&=&\arcsin(\p^2_0)  \nn\\
k^2_3&=&\pi-\arcsin(\p^2_0) \nn\\
k^2_4&=&\pi-\arcsin(\p^2_0-i\uz).
\eea
Again we can derive the S-matrix of this state with the unbound and the
simple bound state. The latter follows from the former upon
application of a relation similar to (\ref{bbsm1}). We find
\be
S_{k\p_0^2}^{u(bb)}=
\frac{\sin k -\p_0^2 -i\frac{u}{2}}
{\sin k -\p_0^2 +i\frac{u}{2}}
\ee
In general there are bound complexes of $2m$ electrons,
($m$-complexes), 
corresponding to a pole in the S-matrix of a $(m-1)$-complex
with a simple bound pair, (a 1-complex). They are parameterized by $m$ complex
numbers $\{\p^m_j$, $j=1\ldots m \}$ in the complex plane:
\be
\p^m_j=\p_0^m+(m+1-2j)i\uv
\ee
corresponding to  $2m$ complex momenta, 
\bea
k^m_1&=&\pi-\arcsin(\p_0^m+im\uv) \nn\\
k^m_2&=&\arcsin(\p_0^m+i(m-2)\uv)  \nn\\
k^m_3&=&\pi-\arcsin(\p_0^m+i(m-2)\uv)\nn\\
\vdots \label{mcomp}\\
k^m_{2m-1}&=&\pi-\arcsin(\p_0^2-i(m-2)\uv) \nn\\
k^m_{2m}&= &\pi-\arcsin(\p_0^2-i(m)\uv)\nn
\eea
In fact, these correspond to the bound complexes 
conjectured by Takahashi \cite{taka}.

\bigskip

We proceed along the same lines as before:
The S-matrix of an
$m$-complex with an unbound particle is,
\be
S_{k\p_0^m}^{u(m)}=
\frac{\sin k -\p_0^m -im\frac{u}{4}}
{\sin k -\p_0^m +im\frac{u}{4}}
\label{sbcub}
\ee
allowing us to  derive the S-matrix of an $m$-complex
with an $n$-complex: 
\be
S_{\p_0^n,\p_0^m}^{(m)(n)}=
\frac{\p_0^m-\p_0^n-|n-m|i\uv}{\p_0^m-\p_0^n+|n-m|i\uv}
\frac{\p_0^m-\p_0^n-(n+m)i\uv}{\p_0^m-\p_0^n+(n+m)i\uv}
\prod^{{\textrm{\scriptsize{min}}}(m,n)-1}_{l=1}
\left(\frac{\p_0^m-\p_0^n-(|n-m|+2l)i\uv}
{\p_0^m-\p_0^n+(|n-m|+2l)i\uv}\right)^2.
\label{smbs}
\ee
If we impose the composite boundary condition on the $m$-complexes,
\be
F(x_1\ldots,x^{(m)}_1,\ldots,x^{(m)}_{2m},\ldots x_N)=F(x_1\ldots 
,x^{(m)}_1+L,\ldots,x^{(m)}_{2m}+L,\ldots x_N)
\label{bsbcm}
\ee
where the $\{x^{(m)}_i\}$ are the coordinates of the members
of the $m$-complex,
 we are led immediately
to
the corresponding eigenvalue of the 
$m$-complex transfer matrix, (recall, all bound states are spin singlets):
\be
\exp(iL\sum_lk^m_l)=
\exp(iL[-2\Re\arcsin(\p_0+mi\uv)+(m+1)\pi])\equiv e^{iq^m(\p_0)L}. 
\ee
In this way we get the full BAC equations,
a generalization of (\ref{rek},\ref{spin},\ref{bsk}):
\be
e^{ik_jL}=\prod_{\d=1}^M\frac{\l_\d-\sin k_j-i\frac{u}{4}}
{\l_\d-\sin k_j+i\frac{u}{4}}
\prod_{(a,n)}S_{\p_0^{(a,n)},k_j}^{(n)u}
\label{rekg}
\ee
\be
\prod_{\d\neq\g}^{M}\frac{\l_\g-\l_\d-i\frac{u}{2}}
{\l_\g-\l_\d+i\frac{u}{2}}
=\prod_{j=1}^{N^u}\frac{\l_\g-\sin k_j-i\frac{u}{4}}
{\l_\g-\sin k_j+i\frac{u}{4}}
\label{sping}
\ee
\be
e^{iq^m({\p^{(b,m)}_0})L}=\prod_{(a,n)\neq(b,m)}S^{(n)(m)}_{\p_0^{(a,n)},
\p_0^{(b,m)}}
\prod_j^{N^u}S_{k_j\p_0^{(b,m)}}^{u(m)}
\label{bskg}
\ee
where the index $(a,n)$ runs over the set of all $n$-complexes present.
These equations were also derived by Takahashi in [\onlinecite{taka}],
within the framework the $k-\L$ string hypothesis. His procedure,
however, involves discarding finite
volume corrections in equations written for finite volume.
Our derivation makes no use of this hypothesis as all
charge bound states are consistently incorporated {\it ab initio}.
\bigskip

To  discuss the nature of  the eigenstates and count them  it is convenient to
consider  (\ref{bskg}) in a logarithmic form,
\be
q^m({\p^{(b,m)}_0})L = \sum_{(a,n)\neq(b,m)}\Th_{nm}\left(\p_0^{(b,m)}
-\p_0^{(a,n)}\right)-2\pi J_m^b
\label{logg}
\ee
where,
\bea
\Th_{nm}(x)&=&\th\left(\frac{x}{|n-m|}\right)+
2\th\left(\frac{x}{|n-m|+2}\right)+\ldots
2\th\left(\frac{x}{|n-m|-2}\right)+
\th\left(\frac{x}{n+m}\right)\ \ \ \ n\neq m\nn\\
&=&2\th\left(\frac{x}{2}\right)+
2\th\left(\frac{x}{4}\right)+\ldots
2\th\left(\frac{x}{2n-2}\right)+
\th\left(\frac{x}{2n}\right)\ \ \ \ \ \ \ \ \ 
\ \ \ \ \ \ \ \ \ \ \ \ \ \ \ \ \ \ \ \ \ \ \ \ \ \ \ \ \ n = m \nn
\eea
with $\th(x)=-2\tan^{-1}(x\vu)$.

Each allowed choice of  quantum numbers $\{J_m^b \}$
uniquely labels the eigenstate, and 
  the allowed ranges  can be deduced \cite{taka}
from eqn(\ref{logg}), (a derivation for the case $m=1$ is given in the next section),
\be
|J_m^b|<\frac{1}{2}(L-N^u-\sum_{n>0}t_{nm}M_n).
\label{phas}
\ee
Here $M_n$ denote the number of $n$-complexes, 
$t_{nm}$ is defined as $t_{nm}=2$min$(n,m)-\d_{nm}$.
These equations are the starting point for counting the number of
states of the model.

We end this subsection by showing that the
 dimensions of the Hilbert spaces for CBC
and PBC are the same. 
The dimension of the Hilbert space with CBC is larger or
equal to the dimension of the space with PBC:
\be
\dim \ {\cal H}_{CBC}\ge\dim \ {\cal H}_{PBC}
\ee
as each vector in ${\cal H}_{PBC}$ lies also in  ${\cal H}_{CBC}$.
On the other hand, there is a one-to-one relation between
 states in ${\cal H}_{CBC}$, not satisfying PBC and states in
${\cal H}_{PBC}$. If one writes a vector in ${\cal H}_{CBC}$
as function of the center-of-mass and relative coordinates
\[
F(x_1,\dots,x_N)=F(X_{c.m.},x_i-x_j) 
\]
one has the injective mapping $\Phi(F)$ onto ${\cal H}_{PBC}$:
\be
\Phi(F)=F(X_{c.m.},[x_i-x_j]\ \ {\textrm{mod}}\ L) 
\ee
because the value of the wavefunction $F(x_1,\ldots,x_N)$
outside the interval $|x_i-x_j|<L/2$ is completely
determined by its value inside. 
It follows
\be
\dim \ {\cal H}_{CBC} = \dim \ {\cal H}_{PBC}
\ee

\subsection{Global Symmetries and the Bethe-Ansatz nature of all Eigenstates}

 We begin by considering
 the global symmetries of the
 model whose 
eigenvalues (partially) label the 
 eigenstates. We shall show, among other things,
 that all the eigenstates 
of the model are of the Bethe Ansatz form. This is in contrast to claims 
by E{\ss}ler, Korepin and Schoutens \cite{ess} that only the highest weight states have the Bethe Ansatz
form while the rest cannot be represented this way.

 First, we
have  the spin $SU(2)$-symmetry, with
\[
\S_z= \frac{1}{2}\sum_i n_{\ua,i}-n_{\da,i},\ \ \  \S^-= \sum_i \psd_{\da,i}\ps_{\ua,i},\ \ \  \S^+= \sum_i \psd_{\ua,i}\ps_{\da,i}
\]

The spin operators commute with the
particle
number operator,
$\N=\sum_i n_{\ua,i}+n_{\da,i}$,
and the application of the spin-lowering
operator $\S^-$ does not lead to a change in
particle number.

The spin highest weight state $|S=S_z = (1/2)(N-2M)\ran$ is defined 
by a solution
$ \{ \l_\g, \ \g=1\ldots M \}$, all $\l_\g$ {\it finite}. To obtain another member  of 
the 
multiplet consider the solution with 
 $M+1$ spin rapidities
$\{ \l_1\ldots \l_M,\l_0 \}$ and $\l_0=\infty$. It
 formally satisfies (\ref{rekg},\ref{sping}), and corresponds 
to the state obtained from $\{\l_1\ldots \l_M\}$
by an application of $\S^-$, since $ B(\l) \ra \S^-/\l$, when $\l \ra \infty$,
in the
framework of the algebraic Bethe ansatz.
Repeated application of $\S^-$ will generate the whole
$SU(2)_{spin}$ multiplet from $\{\l_\g\}$, which
consists of $2S+1$ states, i.e. we will find a
zero wavefunction for more than $2S$ spin
rapidities equal to infinity. A consistent way of defining this process
is provided by considering an anisotropic version of the spin sector.
New complex roots, denote them as $\l^-$ -- roots, are generated with $\pi/\gamma$ as their imaginary part,
where $\gamma$ is the anisotropy parameter. When  $\gamma \ra 0$ isotropy is regained and the $\l^-$ -- roots are sent to infinity in a controllable way, generating
the Bethe Ansatz states that complete the multiplet \cite{kir,kirkir}.

\bigskip
An even simpler  argument holds in
the
case of the charge $SU(2)$ symmetry. It is  defined on a ring
with  $L$ sites ($L$  even), as follows \cite{yang3,affleck},

\be
  \eta_z = \frac{1}{2}(L-\N), \ \  \eta^+=\sum_j (-1)^j \psi_{j \uparrow}^{\dagger}\psi_{j \downarrow}^{\dagger}, \ \  \eta^-=\sum_j (-1)^j \psi_{j \downarrow}
\psi_{j \uparrow}.
\ee

The algebra is consistent with the CBC since the $\eta^{\pm}$ operators
create (destroy) local pairs - see below.
The symmetry is
manifest when one adds a chemical potential term with $\mu=-U/2$
to the hamiltonian:
\be
H_{h.f.}=\sum_{i=-\infty}^{\infty}-t(\psd_{\s,i+1}\ps_{\s,i} +
h.c.)+Un_{\ua,i}n_{\da,i}-\frac{U}{2}\N
\ee
This choice of $\mu$ corresponds to a half-filled system in the
grand canonical formalism.
We have
\be
[H_{h.f.},\eta^{\pm}]=0\ \ \ \ [\N,\eta^{\pm}]=\pm 2\eta^{\pm}
\ee
Clearly, the symmetry generators $\eta^{\pm}$  mix
sectors with different particle number.

In terms of the
BAC equations the $\eta$-symmetry
has a simple explanation: Consider a  given eigenstate of $H$, say $|\ps\ran= |\ps(\{\l_\g\},\{k_j\},\{q_l\})\ran$ with $N$ particles 
characterized by unbound momenta $\{k_j \}$ and bound momenta
$\{ q_l \}$, as well as spin content given by $\{ \l_\g \} $. Acting 
with the operator $\eta^+$
adds to it a 
bound pair with $q=\pi/2$ and crystal momentum $p=\pi$:
\be
\eta^+|\ps(\{\l_\g\},\{k_j\},\{q_l\})\ran
=|\ps'(\{\l_\g\},\{k_j\},\{q_l,\pi/2\})\ran.
\ee
(We have assumed $q_l\neq\pi/2, \forall l$). Note that for a finite system,
namely with CBC for $L<\infty$, the state with $q=\pi/2$
exists  for even $L$.
The state $|\ps'\ran$
is again an eigenstate of $H$, because the S-matrix between the
bound pair with $q=\pi/2$ and all other complexes  is the identity
(total transmission),
-- see (\ref{sbcub}), (\ref{smbs}) and recall: $\p(\pi/2)=\infty$.
The state $|\ps'\ran$ has then $N'=N+2$ particles
and its energy is
\be
H|\ps'\ran=[E(\ps)+U]|\ps'\ran
\label{eigenvh}
\ee
since adding a bound pair at the edge, $q=\pi/2$, corresponds to $\Delta E =U$,
see eqn (\ref{UU}) and subsequent discussion. Thus,
\be
H_{h.f.}|\ps'\ran=E_{h.f.}(\ps)|\ps'\ran.
\ee
$|\ps'\ran$ is degenerate
with $|\ps\ran$ and  the  symmetry
is manifest. We now proceed to  add several
pairs with $q=\pi/2$ to the state $|\ps\ran$. These states have zero width
 - $\xi(\pm \pi/2) =\infty$ - and they
behave as hard-core bosons: The S-matrix among themselves is
$S=-1$,  corresponding to total reflection.
The maximal number of applications
of $\eta^+$ to $|\ps\ran$ is  restricted to
 $L-N$, the number of available
lattice sites. Here $N$ includes bound as well as unbound states - the former
with $q \ne \pm \pi/2$ and thus $\xi \ne \infty$, giving them a finite spread.
For more than $L-N$ applications of $\e^+$  the state is annihilated due to the
Pauli principle. The total number of states 
degenerate under the
$\eta$-symmetry is therefore $L-N+1$. This degeneracy coincides
with the dimension of the $SU(2)_{charge}$ multiplet for
$S_{charge}=S_{z,charge}=  \frac{1}{2}(L-N)   $, which is the eigenvalue of $\eta_z$
applied to $|\ps\ran$. It follows that the Bethe state $|\ps\ran$
is a lowest weight state of this symmetry, and all members of the multiplet have the appropriate Bethe Ansatz form \cite{note}.

Actually, our construction goes further. 
We see from (\ref{eigenvh}), that the $\eta$-symmetry
can be used to group the eigenstates of $H$ into  multiplets
even away from half-filling when the $SU(2)_{charge}$
symmetry is explicitly broken: The energies
of the $L-N+1$ states in the multiplet are equally
spaced with $E_{i+1}-E_{i}=U-2\mu$.

\subsection{Periodic Boundary Conditions  (PBC), and Takahashi's conjecture}

In this section we shall discuss the approach to the bound states
within the usual scheme - imposing periodic boundary
conditions. 

Defining the system on a finite ring $L$ and imposing PBC (a multi-torus
geometry) we have seen above 
that  the parameters
$q$ and $\xi$ of a normalizable eigenstate of $H$ in the two-particle sector
 deviate from the pole (given by eq.
(\ref{xik})) in the
scattering matrix by a term of order $\eul$:
\be
|\cos q|\sinh \xi-\frac{|u|}{4}\sim e^{-\xi L}\sim\eul \ \ {\textrm{for}}
\ \ |q| \ {\textrm{resp.}} \ |q-\pi|\ll \pi/2.
\ee
This deviation is responsible for the instability of this state when
 a third particle is added with real momentum $k$\cite{dan}.
The S-matrix $S^{ub}$ is no longer a pure phase, and 
alters the spin structure of the bound state. There is a finite
(although exponentially small) probability for the bound state to
switch from the spin singlet to the spin triplet state upon passing
through the third particle. As the wavefunction has to be
antisymmetric in configuration space, we have an {\it anti-bound state}
which cannot satisfy periodic boundary conditions and is clearly
forbidden in the infinite volume limit.  That means that PBC 
contradict the local properties of the interaction encoded in the
pole structure of the S-matrix, while the CBC are consistent with it.   
 The coordinate
space of the system with the CBC  is not a multi-torus, but has a more complicated topological structure which, however,
 turns 
into the infinite line for $L\ra\infty$  as does the geometry
of PBC.

One might still look for the analogues of the normalizable
two-particle state (and higher composites)
on a finite ring with PBC -  solutions 
 to the
Lieb-Wu equations (\ref{lw1},\ref{lw2}) with complex momenta. 
It is clear that some of the states must contain complex momenta
because otherwise the subspace containing doubly occupied sites
would be projected out in the $U\ra\infty$ limit \cite{woy}.

Takahashi's $k-\L$ string hypothesis \cite{taka} assumes that
the states containing complex momenta are of the special form
(\ref{stri}) for large $L$, i.e. they become elementary
bound states for infinite volume. Therefore,
using the string hypothesis one obtains 
 BA equations similar to those for CBC
(\ref{rekg},\ref{sping},\ref{bskg}), but containing correction terms. 
In fact, if one drops the
corrections $\sim \eul$ then Takahashi's equations 
do not describe
 PBC but instead CBC for finite $L$. The proof of completeness of the
Bethe ansatz solutions given in [\onlinecite{kor}] uses 
Takahashi's equations for finite $L$
and counts  not the number of states on a finite ring for which 
they are approximate
but on the CBC geometry, where these equations become {\it exact}.
But is Takahashi's hypothesis correct for PBC and finite $L$ when the
terms proportional to $\eul$ do not vanish?

We  argue now, by examining the consequences of this assumption,
that this is not the case and the complete spectrum for PBC is
(in general) not 
given by $k-\L$ string solutions to (\ref{lw1},\ref{lw2}).

We begin by reviewing Takahashi's approach in a simple case with no
spin-strings or higher composites present. We assume $N^e$ 
electrons $2N^b$ of them carrying complex momenta $k_l^{\pm}=q_l\pm i\xi_l$.
In addition we assume $M=N^b+M^u$ real spin rapidities  $\{\L_l, \ \ \l_\g; \ l=1\ldots N^b; \g=1\ldots M^u\  \} $,
$N^b$ of them, $\{\L_l\}$, associated with the complex momenta as follows,

\be
\sin k_l^\pm\equiv \p(q_l)\mp i\chi_l=\L_l\mp i\uv +{\cal{O}}(\eklz).
\label{stri2}
\ee
Plugging this form into (\ref{lw1},\ref{lw2}) one finds,
\bea
e^{ik_jL}=\prod_{\d=1}^{M^u}\frac{\l_\g-\sin k_j-i\frac{u}{4}}
{\l_\g-\sin k_j+i\frac{u}{4}}
\prod_{l=1}^{N^b}\frac{\L_l -\sin k_j -i\frac{u}{4}}
{\L_l -\sin k_j +i\frac{u}{4}}
\label{eqkf}\\
\prod_{\d\neq\g}^{M^u}\frac{\l_\g-\l_\d-i\frac{u}{2}}
{\l_\g-\l_\d+i\frac{u}{2}}
=\prod_{j=1}^{N^u}\frac{\l_\g-\sin k_j-i\frac{u}{4}}
{\l_\g-\sin k_j+i\frac{u}{4}}(1+{\cal E}_\g)
\label{eqsf}\\
\prod_{\d=1}^{M^u}\frac{\L_l-\l_\d-i\frac{u}{2}}
{\L_l-\l_\d+i\frac{u}{2}}
=\prod_{j=1}^{N^u}\frac{\L_l-\sin k_j-i\frac{u}{4}}
{\L_l-\sin k_j+i\frac{u}{4}}(e^{i\ps_l}+{\cal E}_l)
\label{eqsb}\\
e^{2iq_lL}=\prod_{\g=1}^{M^u}\frac{\l_\g-\L_l-i\frac{u}{2}}{
\l_\g-\L_l+i\frac{u}{2}}
\prod_{n\neq l}^{N^b}\frac{\L_n-\L_l-i\frac{u}{2}}{ 
\L_n-\L_l+i\frac{u}{2}}(e^{i\ps_l}+{\cal E}_l).
\label{eqkb}
\eea
Here $\{ {\cal E} \}$ denote terms of order $e^{-\kappa L}$, with
$\kappa\ge |\uv|$. The phase $e^{i\ps_l}$ is defined as  
\be
e^{i\ps_l}=-\frac{\L_l-\sin k_l^+ -i\uv}
{\L_l-\sin k_l^- +i\uv}.
\label{varphi}
\ee
Now the set (\ref{eqkf} - \ref{eqkb}) consists
of $N^u+M^u+2N^b$ coupled algebraic equations for the
variables $\{k_j,\l_\g,q_l,\L_l\}$, i.e. the number
of
variables coincides with the number of equations \cite{note2}. 
But in the $L\ra\infty$ limit, $\L_l$ is not independent from
$q_l$, because of (\ref{stri2}): The $\L_l$ should be eliminated from the set
together with the $N^b$ parameters $\ps_l$, which
do not describe physical properties of the state.
Takahashi did this by substituting eq. (\ref{eqsb}) into (\ref{eqkb})
 after dropping the
$\cal E$-terms. The resulting equations are the 
set  (\ref{rek},\ref{spin},\ref{bsk}). It is clearly
consistent, as it can be derived using BAC. The question arises whether
the finite size correction  terms in
(\ref{eqkf} - \ref{eqkb})
 can spoil the  consistency   for finite $L$. We proceed now to show
that this is the case in general.

Let us assume there is a consistent  solution 
$\{k_j(L),\l_\g(L),q_l(L),\L_l(L),\ps_l(L)\}$
of
(\ref{eqkf} - \ref{eqkb})
for arbitrary (large) $L$.
We define 
the set
$\{k^0_j(L),q^0_l(L),\l^0_\g(L),\L_l^0(L)=\p(q^0_l(L)),\ps^0_l(L)\}$ 
as the
solution
of the zeroth order (in $\cal E$)
terms of (\ref{eqkf} - \ref{eqkb}):
\bea
e^{ik_j^0L}=\prod_{\d=1}^{M^u}\frac{\l^0_\g-\sin k^0_j-i\frac{u}{4}}
{\l^0_\g-\sin k^0_j+i\frac{u}{4}}
\prod_{l=1}^{N^b}\frac{\p(q^0_l) -\sin k^0_j -i\frac{u}{4}}
{\p(q^0_l) -\sin k^0_j +i\frac{u}{4}}
\label{eqkf0}\\
\prod_{\d\neq\g}^{M^u}\frac{\l^0_\g-\l^0_\d-i\frac{u}{2}}
{\l^0_\g-\l^0_\d+i\frac{u}{2}}
=\prod_{j=1}^{N^u}\frac{\l^0_\g-\sin k^0_j-i\frac{u}{4}}
{\l^0_\g-\sin k^0_j+i\frac{u}{4}}
\label{eqsf0}\\
\prod_{\d=1}^{M^u}\frac{\p(q^0_l)-\l^0_\d-i\frac{u}{2}}
{\p(q^0_l)-\l^0_\d+i\frac{u}{2}}
=\prod_{j=1}^{N^u}\frac{\p(q^0_l)-\sin k^0_j-i\frac{u}{4}}
{\p(q^0_l)-\sin k^0_j+i\frac{u}{4}}e^{i\ps^0_l}
\label{eqsb0}\\
e^{2iq^0L}=\prod_{\g=1}^{M^u}\frac{\l^0_\g-\p(q^0_l)-i\frac{u}{2}}{
\l^0_\g-\p(q^0_l)+i\frac{u}{2}}
\prod_{n\neq l}^{N^b}\frac{\p(q^0_n)-\p(q^0_l)-i\frac{u}{2}}{ 
\p(q^0_n)-\p(q^0_l)+i\frac{u}{2}}e^{i\ps^0_l}.
\label{eqkb0}
\eea
The set (\ref{eqkf0} - \ref{eqkb0})
contains again the same number of unknowns as equations and should
give the same solutions as (\ref{rek},\ref{spin},\ref{bsk})
for the parameters $\{k^0_j\},\{\l^0_\g\},\{q^0_l\}$.
In addition it determines the phases $ e^{i\ps_l^0} $.
Because we dropped only the exponentially small correction terms
$\cal E$, the set $\{k^0_j,\l^0_\g,q^0_l,\L_l^0=\p(q^0_l)\}$
deviates from $\{k_j,\l_\g,q_l,\L_l\}$ only 
in quantities of order $\cal E$. Especially, if we choose an
arbitrary $l$ and define $\eps_l=e^{-\xi_lL}$, we can write
\bea
k_j(L)&=& k_j^0(L)+k_j^{(1)}(L)\eps_l \nn\\
\l_g(L)&=&\l^0_g(L)+\l^{(1)}_g(L)\eps_l \nn\\
q_l(L)&=&q^0_l(L)+q^{(1)}_l(L)\eps_l  \label{expans}\\
\L_l(L)&=&\p(q^0_l(L)) + \L^{(1)}_l(L)\eps_l 
\nn \\
\chi_l(L)&=&\uv +\chi_l^{(1)}(L)\eps_l.
\nn
\eea
We will now show  that the coefficient $\L^{(1)}_l$
in (\ref{expans})
is determined through two independent equations, leading to
an over-constraint.
 
One set of equations to determine  $\L^{(1)}_l$ is obtained as follows:
dividing 
(\ref{lw1}) for $k_l^+$ with (\ref{lw1}) for $k_l^-$
we get the following equation:
\be
e^{-2\xi_lL}=
\frac{(\L_l-\p(q_l))^2+(\xi_l-\uv)^2}{(\L_l-\p(q_l))^2+(\xi_l+\uv)^2}
\prod_\g^{M^u}
\frac{(\l_\g-\p(q_l))^2+(\xi_l-\uv)^2}{(\l_\g-\p(q_l))^2+(\xi_l+\uv)^2}
\prod_{n\neq l}^{N^b}
\frac{(\L_n-\p(q_l))^2+(\xi_l-\uv)^2}{(\L_n-\p(q_l))^2+(\xi_l+\uv)^2}.
\ee
which leads to,
\be
1=\frac{4}{u^2}\left(1-
\left[\frac{e^{i\ps_l^0}-1}{e^{i\ps_l^0}+1}\right]^2\right)
(\chi_l^{(1)})^2f(\{\l_\g^0,q_n^0\})
\label{fior}
\ee
with  $f$
some function of $\{\l_\g^0,q_n^0\}$. This equation determines
$\chi_l^{(1)}$ 
as function of the zeroth order variables
$\{\l_\g^0,q_n^0,\ps_l^0\}$, therefore also the
coefficient $\L^{(1)}_l$ in (\ref{expans}), (using
(\ref{varphi})),
up to terms which are by themselves exponentially small. 
On the other hand, we obtain $\L^{(1)}_l$
directly from  
eq. (\ref{eqsb}), by looking at the
$\eps_l$-correction.  These two determinations of the
coefficient $\L^{(1)}_l$  are independent, as the former
is derived from (\ref{lw1}) and the latter from (\ref{lw2}).
This
indicates that the
set (\ref{eqkf} - \ref{eqkb}) is
over-constrained
in the first order finite volume correction to the thermodynamic limit.
This effect can be studied in the three particle case by an
explicit construction of the BA wave function assuming
 the $k-\L$ string hypothesis. An  over-determination
of the parameters in expressions of order $\cal E$ is found\cite{dan}. 
We conclude that the $k-\L$ string hypothesis
does not correspond to an actual solution of the BA equations for
sufficiently large but finite $L$. The spectrum of the model on a finite ring
is not in analytic one-to-one correspondence with the spectrum
on the infinite line. On the other hand, the
CBC - spectrum develops smoothly into the infinite volume limit,
probably because it already contains an additional symmetry. This symmetry
is presumably destroyed by the PBC and appears in this case only in the
 infinite volume limit. 

Another way to observe the over-determination of (\ref{lw1},\ref{lw2})
for finite volume is based on the algebraic Bethe ansatz.

We  show that 
equations (\ref{eqsb}),  which do not contain $L$ explicitly,
become redundant for  $L\ra\infty$, leading to superfluous
constraints on the parameters: assume a pair of  complex
 conjugated momenta $k^+, k^-$ related to the
spin momentum $\L$ by (\ref{stri}).
An eigenstate  of the corresponding inhomogeneous
transfer matrix $Z(\mu)$ with (arbitrary) spectral parameter $\mu$ 
 then has the form (see e.g. [\onlinecite{and}]): $
|\ps\ran = \prod_{\g=1}^{M^u}B(\l_\g)B(\L)|\o\ran$
for  total spin $S=\frac{1}{2}(N^u-2M^u)$, 
where the $M^u$ creation operators $B(\l_\g)$, acting on the
ferromagnetic vacuum $|\o\ran$, create $M^u$ $\da$-spins, and
$B(\L)$ creates the $\da$-spin of the bound state. 
By explicit calculation we find that $B(\L)$ acting on $|\o\ran$
 diverges as $\eulp$
for $L\ra\infty$, whereas the $B$'s not associated with the
complex pair do not diverge. Normalizing $|\ps\ran$ by
multiplication with $\eul$ yields
accordingly an  exponential suppression of vectors of the form
$\prod_{\d}B(\l_\d)|\o\ran$ which {\it do not} contain
$B(\L)$ among the $B(\l_\d)$. Now the equation (\ref{eqsb}) for $\L$
is necessary to cancel an ``unwanted term'' in the eigenvalue
equation for 
$|\ps\ran$:
\be
(Z(\mu)-E(\mu))|\ps\ran=  
\sum_{\g}^{M^u}\a_\g\prod^{M^u}_{\d\neq\g}B(\l_\d)B(\L)B(\mu)|\o\ran
+\a_0\prod_{\g}^{M^u}B(\l_\g)B(\mu)|\o\ran. 
\ee
It ensures, in particular, that $\a_0=0$. But, as argued previously,
the vector $\prod_{\g} B(\l_\g)B(\mu)|\o\ran$ is
projected out in the infinite volume limit (exponentially suppressed
for
finite $L$), and therefore (\ref{eqsb}) is not necessary for $|\ps\ran$
to be an eigenvector of $Z(\mu)$ for $L\ra\infty$. That means that only a 
subset of all states allowed in the infinite volume can
be  generated by the $L\ra\infty$ limit of string solutions to 
(\ref{lw1},\ref{lw2}) for
finite $L$.

In conclusion we find that the $k-\L$ string hypothesis represents an
over-constrained ansatz for solutions of BA equations for periodic
boundary conditions, having in general no solutions for large
but finite volume $L$.

\section{ gapped excitations}

In this section we will compute the simplest gapped excitations
above the ground state
for the repulsive and the attractive case. In the former case this
amounts to the formation of a bound pair  above the sea of unbound 
particles, in the latter case one of the
bound pairs forming the ground state is broken and the resulting
excitation has nonzero spin. Both excitations are characterized by
a gap of order $U$ and are accompanied, when the number of electrons is held fixed, by two gapless excitations: two holons in repulsive or two dressed electrons in the attractive case. 

 The  range of the gapped excitation momentum, $p^{bs}$, depends on the 
filling $n$,  $-\pi(1-n) \le p^{bs} \le\pi(1-n)$. At half filling it becomes
 therefore a non dynamic mode. We present the dispersion of the gapped mode
for various fillings and interaction strengths.

\subsection{ The repulsive case}
 A single bound pair above the ground state of the repulsive Hubbard
model can be created without changing the particle number $N$ by
placing two holes in the sea of the charge quantum numbers.
$N^u=N-2$ and $M=\frac{N}{2}-1$ in the notation given above. Equations
(\ref{rek},\ref{spin},\ref{bsk})  then read \cite{haldane},
\bea
Lk_j = 2\pi n_j+\sum_{\d=1}^M\Th_1(\sin k_j-\l_\d)
+\Th_1(\sin k_j -\p(q))
\label{rek2} \\
\sum_{j=1}^{N-2}\Th_1(\sin k_j-\l_\g)=\sum_{\d=1}^{M}\Th_2(\l_\d-\l_\g)
+2\pi I_\g
\label{spin2}\\
2qL=\sum_{j=1}^{N-2}\Th_1(\p(q)-\sin k_j) +2\pi J
\label{bsk2}
\eea
where $\Th_n(x)=\th(x/n)$.
The range for the quantum numbers $\{n_j\}$ is:  $-N/2\le n_j\le N/2-1$ for
$M=N/2-1$ even and $-(N-1)/2\le n_j\le (N-1)/2$ if $M$ is odd.
The $\{I_\g\}$ range between $-((N-2)-M-1)/2$ and $+((N-2)-M-1)/2$.
The $\{n_j\}$ sequence contains two holes as the actual number of free
$k$-momenta is $N-2$. The $\{I_\g\}$ sequence does not contain holes in
the
absence of spin excitations.
 $J$, the quantum number associated with
the bound state, is an  integer if $N$ is even and a half-odd
integer if $N$ is odd. We assume $N$ even in the following.
To find the limiting values for $J$, we consider the boundaries of the
allowed range for $q$: $\pi/2< q <\pi$, (resp. $-\pi<
q<-\pi/2$). We may treat the range for $q$ as connected by
shifting $-\pi/2$ to $3\pi/2$.
Setting $q=\pi/2$:
\be
\pi L = 2\pi J^- -(N-2)\pi
\label{ran1}
\ee
as $\p(\pi/2)=\infty$. For $q=3\pi/2$ we have $\p(q)=-\infty$. Thus,
\be
3\pi L= 2\pi J^+ +(N-2)\pi.
\label{ran2}
\ee

Hence,
\be
L-\frac{N}{2}> J > \frac{N}{2}
\ee
At half-filling: $(N/2)\le J\le (N/2)$, which leads to maximal restriction
of the phase space for the bound state. 
In the infinite volume limit which is treated here, the phase space
for
$J$ vanishes at half-filling. Recall that
this is the  case
for spin singlet excitations where the string parameter is given in terms
of the
hole parameters \cite{and}. The Lieb-Wu equations studied in 
[\onlinecite{woy}] lead to a similar rigid relation between $q$,
$k_1$ and $k_2$, namely $\p(q)=\frac{1}{2}(\sin k_1+\sin k_2)$.
While this rigidity holds  at half-filling it is physically meaningless,
as the physical momentum $p^{bs}$ becomes {\it independent}
of $q$ at this point (see below).

We  proceed now in the usual manner \cite{coll}, 
using the notation of [\onlinecite{and}]. One introduces the functions
$\{ \r(k),\s(\l) \}$ to describe the densities of the $\{k\}$ and $\{\l\}$
solutions respectively. The ground state densities $\{
\r_0(k),\s_0(\l) \}$ are solutions of the appropriate integral
equations with no holes in the distribution of the $k_j$,
while the state under consideration now is given in terms
of$\{ \r_b(k),\s_b(\l) \}$.
 The integral equations for
these densities 
read,
\begin{displaymath}
\r_b(k)+\frac{1}{L}\d(k-k_1))+\frac{1}{L}\d(k-k_2)= 
\end{displaymath}
\be
\frac{1}{2\pi}+\cos k\int \rd\l\s_b(\l)K_1(\sin k-\l)+\frac{1}{L}
\cos k K_1(\sin k-\p(q))
\label{inr}
\ee
\be
\s_b(\l)=\int_{-Q}^Q \rd k\r_b(k) K_1(\sin k -\l)-\int \rd\l'\s_b(\l')
K_2(\l-\l')
\label{ins}
\ee
where $k_1, k_2$ denote the positions of the holes, and $K_1, K_2$ as well as $R,\K^Q$ (see below) are kernels of integral operators defined as in [\onlinecite{and}]. The parameter $Q$ defines the range of $k$ and is given by,
\be
\bear{ll}
\int_{-Q_0}^{Q_0}\rd k\r^{Q_0}_0 &= \frac{N}{L}\\
\int_{-Q}^{Q}\rd k\r^{Q} &= \frac{N-2}{L}
\eear
\ee
it depends therefore
implicitly on $k_1,k_2$ and $q$.
The equation
for $q$ reads,
\be
2q = \frac{2\pi}{L}J + \int_{-Q}^{Q}\rd k\r(k)\Th_1(\p(q)-\sin k).
\ee
It is convenient to introduce $\r'_1(k),\s_1(\l)$,
\be
\bear{ll}
\r(k)&=\r_0^Q(k)+\frac{1}{L}\r'_1(k) -\d(k-k_1)-\d(k-k_2) \\
\s(\l)&=\s_0(\l)+\frac{1}{L}\s_1(\l)
\eear
\ee
these correspond to the density changes induced by the excitations with respect to the density $\r^Q_0$. 
The Fourier transform of the 
spin density $\s_1(\l)$ reads in terms of $\r_1'(k)$,
\be
\tilde{\s}_1(p)=
\frac{1}{2}\int_Q^Q\r_1'(k)\sech (\uv p)e^{-ip\sin k}
-\frac{1}{2}\sech (\uv p)[e^{-ip\sin k_1}+e^{-ip\sin k_2}].
\ee
It is possible to split $\r_1'(k)$ into three parts
\be
\r_1'(k)=\r^c_1(k,k_1)+\r^c_2(k,k_2)+\r^{bs}(k,q)
\ee
such that
\be
\K^Q[\r^c_j](k)=-\cos k\vu R\left(\vu(\sin k-\sin k_j)\right)
\label{inteqj}
\ee
for $j=1,2$ and
\be
\K^Q[\r^{bs}](k)=\cos kK_1(\sin k -\p(q))
\label{inteqbs}
\ee
The excitation energy $\Delta E=E(Q)-E_0(Q_0)$ is defined via
\be
\bear{ll}
E_0(Q)&=-2tL\int_{-Q}^Q\rd k\r^Q_0(k)\cos k\\
E(Q)&=-2tL\int_{-Q}^Q\rd k\r(k)\cos k
\eear
\ee
With the usual definition of the chemical potential $\mu$
\be
\mu(Q)=\left(\frac{\partial E_0}{\partial N_0}\right)=
\left(\frac{\partial E_0(Q)}{\partial Q}\right)
\left(\frac{\partial N_0(Q)}{\partial Q}\right)^{-1}
\ee
and
\be
N_0(Q)=L\int_{-Q}^Q\rd k \r^Q_0
\ee
It follows
\be
 \Delta E = -2t\int_{-Q}^Q\rd k \r_1'\cos k
+2t[\cos k_1 +\cos k_2] - \mu(Q_0)\int_{-Q}^Q \rd k \r_1'
\ee
$\mu$ is independent of $k_1,k_2$ and $q$ and given by
Coll's formula \cite{coll}:
\be
-\frac{\mu(Q_0)}{2t}=
\frac{\cos(Q_0)-\int_{-Q_0}^{Q_0}\rd k \cos k\r^c(k,Q_0)}
{1-\int_{-Q_0}^{Q_0}\rd k\r^c(k,Q_0)}
\label{chempot}
\ee
The excitation energy is a sum of three parts
\be
\Delta E = E^h_1 +E^h_2 +E^{bs}
\ee
Each of them consists of a direct, a backflow and a 
ground state contribution:
\be
E^h_j(k_j) = 2t\cos k_j -2t\int_{-Q}^{Q}\rd k\r_c(k,k_j)\cos k
-\mu\int_{-Q}^{Q}\rd k\r_c(k,k_j)
\ee
The hole-energy  goes to $-\mu$ for $k_j\ra Q$.
The bound state energy reads
\be
E^{bs}(q) = -4t\cos q\cosh(\xi(q)) -2t\int_{-Q}^Q\rd k \r^{bs}\cos k
-\mu \int_{-Q}^Q\rd k \r^{bs}.
\label{bsen}
\ee

We proceed to discuss the energy dispersion. We shall do it for any filling.
Consider the half filled case:
using the identity $
-4t\cos q\cosh(\xi(q))=U + 2t\int_{-\pi}^\pi\rd k \cos^2kK_1(\sin k
-\p(q))
$ we see
$E^{bs}(q)-U$ vanishes identically at half-filling and the dispersion
of the excitation depends only on the hole part apart from the
gap $U$. 

 Away from half-filling the bound state contribution
 becomes an independent
excitation with its own dispersion.  The momentum has contributions
from the holes and from the bound state,  given by,
\be
\Delta P = P-P_0 = -p^h_1-p^h_2 +p^{bs}.
\ee

The hole-momenta $\frac{2\pi}{L}n_j$ read
\be
p^h_j=\int_0^{k_j}\rd k\r_0(k)
\ee
(which leads to the identification of the point $k_j=\pm Q$ with the
charge Fermi-momentum $\pm k_F=\pm\pi(N/L)$), while
for $p^{bs}$ we find
\be
p^{bs}(q)=2q-\int_{-Q}^Q\rd k\r_0(k)\Th_1(\p(q)-\sin k).
\label{bsmo}
\ee
As $q$ runs over the allowed range 
$\pi/2<q<\pi$ (resp. $-\pi<q<-\pi/2$) , $p^{bs}$ varies between $\pi+k_F$ and
$2\pi$
(resp. $-2\pi<p^{bs}<-\pi-k_F$). This corresponds to a
symmetric band around zero between $-|\pi-k_F|$ and $\pi-k_F$.
It follows that at half-filling $p^{bs}\equiv 0$, i.e. the
phase volume vanishes. The physical meaning of the shrinking to zero of parameter range 
at half filling has a simple  interpretation: the bound state has no room to propagate.
Although the bound state parameter $q$ is fixed in this case, in
terms
of the hole parameters $k_1$ and $k_2$ \cite{woy}, this has no physical 
meaning:
Energy and momentum,
the physical parameters of the excitation, are {\it independent} of
the
unphysical parameter $q$, namely $E^{bs}\equiv U$ and $p^{bs}\equiv
0$. 
This redundancy of the parameter $q$ at half-filling has led to
some misconceptions in the literature \cite{note3}.
  
Away from half-filling, the
dispersion of the (always gapped) excitation is given parametrically by
(\ref{bsen}) and (\ref{bsmo}) for fixed hole-momenta $k_1$ and
$k_2$. We solved (\ref{inteqj},\ref{inteqbs})
numerically for $t=1$ and  two values of the interaction strength, $U= 0.5$
 and $U=2$.
Figures 1 and 2 show the dispersion of the bound state energy
$E^{bs}(p^{bs})$ for $U=2$ and $U=0.5$ respectively and for
 different
values of the density $n$. Figures 3 and 4 show the dispersion curves for the same values of $U$, but for densities very close to half filling.

 We notice that beyond a critical filling $n_c(U)$  the bound 
state energy dips below $U$. The critical value depends on $U$ and decreases
with increasing $U$.  We are 
currently exploring whether the gap
between the top of the holon and spinon bands 
and the bottom of the upper Hubbard band can actually close for some value
of $U$.

\begin{figure}
\center
\epsfig{width=0.7\textwidth,file=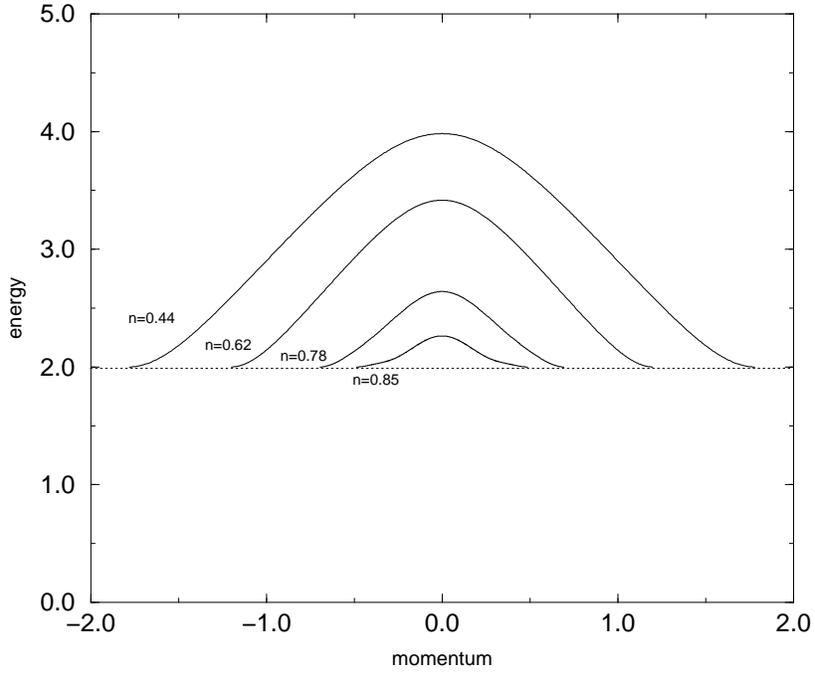}
\caption{The band of the bound state excitation for $U=2$,
and densities $n=0.44, 0.62, 0.78$ and $0.85$, respectively.}
\end{figure}
\begin{figure}
\center
\epsfig{width=0.7\textwidth,file=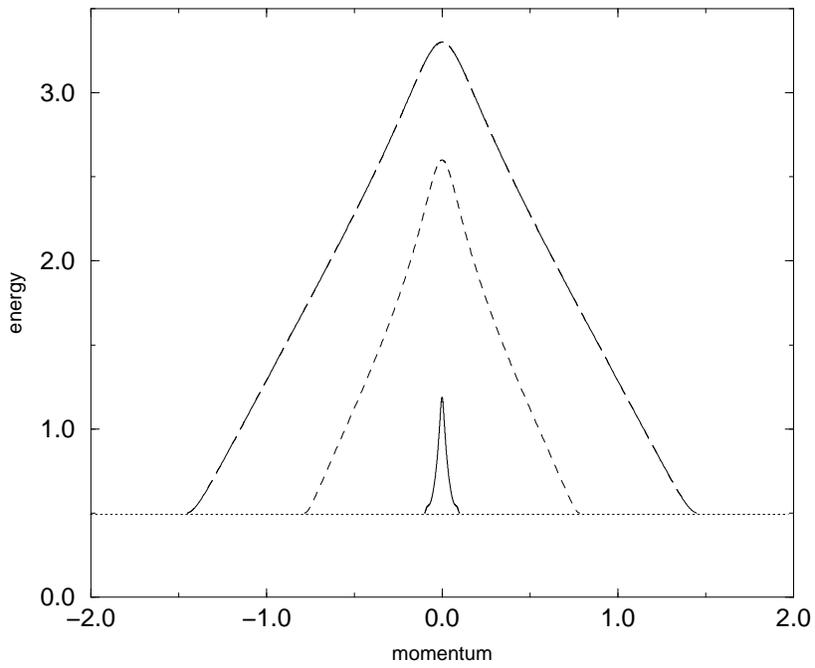}
\caption{The band for $U=0.5$ and three values for the density,
$n=0.53$ (long-dashed), $n=0.75$ (dashed), and $n=0.96$ (solid line).}
\end{figure}
\begin{figure}
\center
\epsfig{width=0.7\textwidth,file=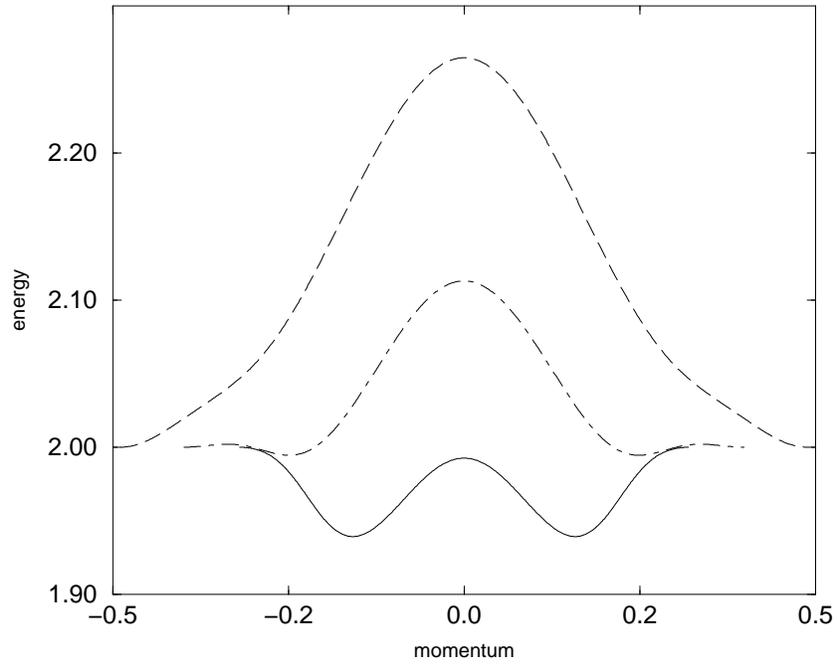}
\caption{The dispersion for $U=2$ and $n=0.84$ (long-dashed),
$n=0.87$ (dot-dashed) and $n=0.89$ (solid line).
The critical density $n_c\sim 0.86$.}
\end{figure}
\begin{figure}
\center
\epsfig{width=0.7\textwidth,file=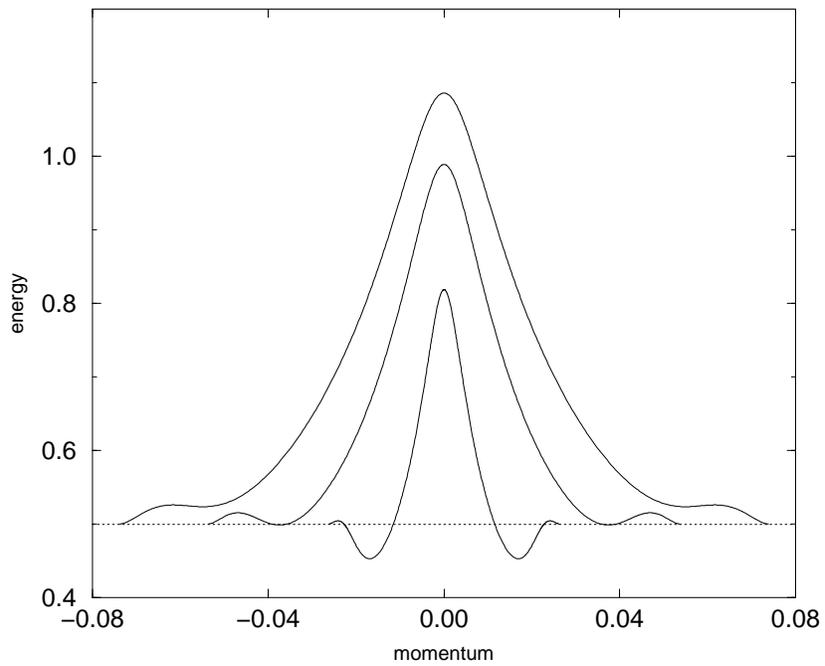}
\caption{The dispersion for $U=0.5$ and three densities close to half
filling:
$n=0.97$, $0.98$ and $0.99$.
The critical density $n_c= 0.98$.}
\end{figure}

\subsection{The attractive case}
In the attractive case $U<0$ and we expect the ground state to 
consist solely of bound pairs (for even $N$), being a total spin
singlet
with $N^b=N/2$ and $N^u=M^u=0$. We find accordingly,
\be
e^{2iq_l}=
\prod_{n\neq l}^{N^b}\frac{\p(q_n)-\p(q_l)-i\frac{u}{2}}{ 
\p(q_n)-\p(q_l)+i\frac{u}{2}}
\label{bsk1a}
\ee
a reduced form of (\ref{bsk}). We proceed in
analogy with the treatment in section III A.
The set of the $q_l$ lies in the interval $-\frac{\pi}{2}< q_l
< \frac{\pi}{2}$.
Taking the logarithm of (\ref{bsk1a})
\be
2q_lL=2\pi J_l-\sum_{n\neq l}^{N^b}\Th_2(\p(q_l)-\p(q_n))
\ee
The 
range for the quantum number
$J_l$ , corresponding to the range for $q_l$ above, is
\be
-(L-N^b-1)/2 < J_l < (L-N^b-1)/2
\label{rangej}
\ee
i.e. $J_l$ is integer (half-odd-integer) if $L-N^b$ is odd (even).
At half-filling $N=L$, the  $J_l$'s are filling
all the slots allowed by (\ref{rangej}). 
It is convenient to change variables from $q_l$ to $\p(q_l)$, in
defining the density, $L\s(\p)=\rd J/\rd \p(q)$
which leads to the ground state integral equation,
\be
\Li^{B_0}[\s](\p) = \frac{1}{\pi}F(\p)
\label{gsea}
\ee
with $
\Li^{B_0}[\s](\p) = \s(\p) + \int_{-B_0}^{B_0} \rd\p
K_2(\p-\p')\s(\p') $
and the inhomogeneous term, $
F(\p)=\rd q(\p)/\rd\p=\Re[1/\sqrt{1-(\p+i\uv)^2}]$.
As observed in [\onlinecite{and}], operator $\Li^{B}$ plays the same
role as $\K^Q$ in the repulsive case. The integration limit
$B_0$ is determined by $
\int_{-B_0}^{B_0}\s(\p) = N^b/L$.
At half-filling the r.h.s. is $1/2$,    
 allowing us to deduce from that $B_0=\infty$, and
the equation  can be solved via Fourier transformation 
\cite{and}.

We now consider the simplest gapped excitations above this ground state.
It involves 
 pair-breaking, and is therefore a spin excitation. 

We consider first the
triplet: it is 
created by removing one bound pair and adding two free particles in a
triplet spin state.
We have, 
\bea
2q_lL&=&2\pi J_l-\sum_{n\neq l}^{N^b-1}\Th_2(\p_l-\P_n)-
\Th_1(\p_l-\sin k_1)-\Th_1(\p_l-\sin k_2)
\label{trip1}\\
k_jL&=&2\pi n_j - \sum_{l}^{N^b-1}\Th_1(\sin k_j-\p_l)
\label{trip2}
\eea
for $j=1,2$. 
The total spin is $S=1$, $N^b\ra N^b-1$, $N^u=2$,
$M^u=0$. 

At half-filling the number of available slots is reduced
by one (see (\ref{rangej})), therefore no hole opens in the $J$-
sequence. The parameter of the excitation are just the two 
momenta of the free particles $k_1$ and $k_2$.
Away from half-filling $N=2N^b<L$ and not all allowed
slots are occupied in the ground state. Now, $B_0<\infty$,
the $J_l$'s are distributed symmetrically
around zero and
$|q_{max}| < \pi/2$. That means, we can create a hole in the
$J$- sequence at position $J_h$, which corresponds to a $|\p_h|
<B_0$ although the number of allowed slots decreases by one.
This $J_h$ is the third parameter of the spin-triplet excitation
and corresponds to the bound state parameter in the repulsive case.
We proceed by introducing the densities, $
\s(\p)=\s_0^{B}(\p)+\frac{1}{L}\s_1(\p)$ and $
\s_1(\p)=\s_1'(\p)-\d(\p-\p_h)$ where $\s_0^{B}(\p)$, being the ground state
distribution with Fermi level $B$ rather than $B^0$,is determined by the normalization condition $
\int_{-B}^{B}\s(\p) = (N^b-1)/L$.

Expressing the smooth density $\s_1'(\p)$ as a sum of three terms, $
\s_1'(\p) = \s_c^1(\p) + \s_c^2(\p) + \s_h(\p)$ we find,
\be
\Li^B[\s_c^j](\p) = -K_1(\p-\sin k_j)
\label{sigfr}
\ee
for $j=1,2$ and
\be
\Li^B[\s_h(\p,\p_h)](\p) = K_2(\p-\p_h).
\label{sigho}
\ee
Having solved (\ref{sigfr},\ref{sigho}) for the densities
$\s_c^j,\s_h$,
we compute the excitation energy.
The total excitation energy consists of three terms,
\be
\Delta E = E_1+E_2+E_h = -2t\sum_{j=1}^2\cos k_j-\E(\p_h)+\int_{-B_0}^{B_0}
\E(\p)\s_1'(\p)
-\mu^\s(B_0)\int_{B_0}^{B_0} \s_1'(\p),
\label{toen}
\ee
associated with the two unbound electrons and the 
{\it independent} hole respectively,
\bea
E_j &=& -2t\cos k_j +\int_{-B_0}^{B_0}\s_c^j(\p)[\E(\p)-\mu^{\s}] \label{efree}\\
E_h &=& -\E(\p_h) +\int_{-B_0}^{B_0}\s_h(\p,\p_h)[\E(\p)-\mu^\s].
\label{ebound}
\eea
$\mu^\s$ is the chemical potential and accounts for the shift of
the Fermi momentum $q(B)$ through the excitation. It is defined
with respect to the number of bound pairs, $\mu^{\s} = \rd E_0/\rd N^b$,
and given by an analog to
(\ref{chempot}):
\be
\mu^\s(B_0)=\frac{\E(B_0)-
\int_{-B_0}^{B_0}\E(\p)\s_h(\p,B_0)}{1-\int_{-B_0}^{B_0}\s_h(\p,B_0)}.
\ee
$\E(\p)$ denotes the bound state energy function,
\be
\E(\p) = -4t\Re\sqrt{1-(\p+i\uv)^2} \le -|U|.
\ee

At half-filling $\s_h(\p,\infty)\equiv 0$ and
$\mu^{\s}(\infty)=-|U|$, as expected \cite{and}.

The momenta $p_j;\  p_h$,  associated with the dressed electrons $j=1,2$
 and the hole respectively, are given by, 
\bea
p_h&=&\int_0^{\p_h}\s_0^{B_0}(\p) 
\label{pbound} \\
p_j &=& k_j + \int_{-B_0}^{B_0}\rd\p\Th_1(\sin k_j-\p).
\label{pfree}
\eea

Removing the bound state at the Fermi level $B$ we have,
\be
p_h(B)=\pi\frac{N^b}{L}= \frac{\pi}{2}n
\ee
which identifies the Fermi momentum in the attractive case,
$k_F^{att}=\frac{\pi}{2}n$. Note also that
the dressed momenta $p_j$ of the two unbound electrons deviate from their
free values $k_j$.
The hole contribution to energy and momentum, given by 
$E_h$ and $p_h$ in  (\ref{ebound}) and (\ref{pbound}),  play the
same role as $E^{bs}$ and $p^{bs}$ in section III A.

Now consider the singlet excitation. We break a pair and put the two
electrons into a spin singlet state.
Again $N^b$ is reduced by one, leading to an additional
degree of freedom away from half-filling. $N^u=2$, but $M^u=1$ and
$S=0$, the total spin is not changed. 
We have one  $\l$ parameter for the unbound electrons.
Equation (\ref{spin}) reads then,
\be
1=\frac{\l-\sin k_1+i\frac{|u|}{4}}{\l-\sin k_1-i\frac{|u|}{4}}
\frac{\l-\sin k_2+i\frac{|u|}{4}}{\l-\sin k_2-i\frac{|u|}{4}} 
\ee
which leads to the familiar form
\be
\l=\frac{1}{2}(\sin k_1 +\sin k_2).
\label{lasing}
\ee
Eq. (\ref{trip1}) remains valid in the singlet case while
eq. (\ref{trip2}) becomes
\be
k_jL=2\pi n_j - \sum_{l}^{N^b-1}\Th_1(\sin k_j-\p_l) -\Th_1(\sin k_j-\l)
\label{sing2}
\ee
The  effect of this modification is to change  the relative
phase shift of the particles,
\be
\d^{singlet}=\d^{triplet}-\Th_2(\sin k_1-\sin k_2)
\ee
as expected on general grounds\cite{and}.

\bigskip

We have analyzed the simplest gapped 
excitations. In the repulsive case, when the number of electrons is
held fixed, the spectrum consists of two gapless holons, as well as 
a gapful singlet residing in the upper Hubbard band. We have calculated its
dispersion for various values of the filling and interaction strength, see
figure (1,2,3). The  
gapped excitations are independent of the concomitant
holon excitations away from half-filling.
A dual picture holds for the attractive case.

 The upper Hubbard excitation we discussed was based on a
two-particle bound state; other excitations emerge when
 higher composites  are studied, and 
can be identified with elementary $m$-complexes, see eq.(\ref{mcomp}).
 These  have higher gaps with respect to
the ground state and constitute the upper bands.

\section{Summary}

We have studied in this article the gapful excitations of the
one-dimensional Hubbard model, constituting
the upper Hubbard band. We have introduced a simple and intuitive construction
to incorporate complex momenta solutions that correspond to these states.
 Our analysis has the following results:

1)\ \ In infinite volume  states with complex momenta
are (elementary) bound states of the hamiltonian, formed of
any (even) number of electrons. These states are spin
singlets and live in the upper Hubbard band for $U>0$ and the lower band for
$U<0$. They are renormalized by interacting
 with other particles, bound or unbound.

2)\ \ These states exist  for finite volume if
Composite Boundary Conditions (CBC) are introduced.
Their existence can be proven exactly
without postulating a string-hypothesis, via
 the Bethe Ansatz for Composites (BAC) approach. We have shown that
all states can be obtained within this scheme.

3)\ \ We clarified the physical interpretation
 of the $SU(2)_{charge}$ symmetry
in the framework of BAC and have shown that it corresponds to the addition of
completely local bound states.  A simple proof for
the corresponding lowest weight property of the Bethe states follows.

4)\ \ We have argued that the states with complex momenta 
{\it cannot} have the string form for large but finite volume and
periodic boundary conditions (PBC). That means that the $k-\L$ string
hypothesis may give correct results in the thermodynamic limit but
fails for finite volume and PBC.

5)\ \ The simplest  gapped  elementary excitations 
involve a multi-particle bound state. When the number of electrons is held
 fixed the presence of such a state is accompanied by the appearance of
at least two holons in the repulsive case or two dressed electrons in the
attractive case. The gapped excitation is an independent mode only
away from half filling, becoming a non-dispersive gap at half filling.

\bigskip

We conclude with two conjectures. The first is related to the possibility
of the BAC construction, which follows from the composite boundary
conditions:
we conjecture that the CBC incorporate a symmetry algebra which should
be
similar to the Yangian symmetry
found up to now only in infinite systems.

Furthermore, we conjecture that for a finite system with periodic
boundary
conditions the Hubbard model is no longer integrable in the strong
sense;
that means the wavefunctions 
in the $N$-particle sector with spin $S$ cannot be parameterized by the set
$\{k_j,\l_\g\}$
of $N$ momenta and $N/2-S$ spin rapidities alone, if some of the
momenta are complex.

\bigskip

We wish to thank A. Ruckenstein, T. Kopp,  R. Fresard, C. Bolech, A. Jerez 
and P. Zinn - Justin
for  carefully reading  the manuscript
and for their insightful comments.

\subsection{Appendix - Derivation of eq.(\ref{sub})}
This appendix
contains the
detailed derivation
of (\ref{sub}), thereby showing the stability of the two-particle
bound state if it satisfies the pole condition (\ref{xik}).
We define
$S^{ub}_{1(23)}$ by,
\be
A^{a_1a_2a_3}_{[231]}=S^{ub}_{1(23)}(k,q^-,q^+)A^{a_1'a_2'a_3'}_{[123]}
\ee
i.e. it takes
particle 1 with the
real momentum $k$
from the left side
of the bound pair
to the right side,
without changing
their momenta. In
both the initial and
final amplitudes, we
have particle 2 on
the left of particle
3, which means
$k_2=q^-$ and
$k_3=q^+$.
Therefore,
$S^{ub}_{1(23)}$ is
given as the product,
\be
S^{ub}_{1(23)}(k,q^-,q^+)=S^{uu}_{13}(k,q^+)S^{uu}_{12}(k,q^-)
\ee
which is equation
(\ref{frb}). To
proceed we write
$S^{uu}(k_a,k_b)$ in
the form
\be
S^{uu}_{ab}(k_a,k_b)=\frac{1}{2}(1+s_{ab})\id
+\frac{1}{2}(1-s_{ab})P_{ab}
\ee
where $P_{ab}$ is
the spin exchange
operator between
particles $a$ and
$b$.
The phase $s_{ab}$
depends on $k_a$ and
$k_b$ as,
\be
s_{ab}=\frac{\sin
k_a-\sin k_b -i\uz}{
\sin k_a-\sin k_b +i\uz}.
\ee
We write
\be
s_- = \frac{\sin
k-\sin q_- -i\uz}{
\sin k-\sin q_- +i\uz}\ \
\ \ 
s_+ = \frac{\sin
k-\sin q_+ -i\uz}{
\sin k-\sin q_+ +i\uz}.
\ee
Then
\bea
S^{ub}_{1(23)}&=&
\left(\frac{1}{2}(1+s_+)+\frac{1}{2}(1-s_+)P_{13}\right)
\left(\frac{1}{2}(1+s_-)+\frac{1}{2}(1-s_-)P_{12}\right)\nn\\
 &=&
\frac{1}{4}\left[(1+s_+)(1+s_-)+(1+s_-)(1+s_+)P_{13}\right.\label{equa1}\\
&\ph{=}& \left.+(1+s_+)(1+s_-)P_{12}
+(1-s_+)(1-s_-)P_{13}P_{12}\right].\nn
\eea
Now we use the fact
that the spin space
of the three
particles is
restricted: 2 and 3
are in a mutual
singlet in region
$[123]$. The spin
state space in this
region is therefore
$V_A=V_1\otimes
V^{singlet}_{23}$
and a two-dimensional
subspace of the
eight-dimensional
spin space of the
three particles.
Under this
condition
$P_{23}A_{[123]}=-A_{[123]}$.
It follows then
$P_{13}P_{12}=P_{12}P_{23}=-P_{12}$
if acting on
$V_A$. A further
identity valid if
the operators act on
$V_A$ reads $P_{12}=\id-P_{13}$. 
Note that
these operators do
not leave $V_A$
invariant. 
In
general, therefore,
the singlet state of
particles 2 and 3
will be destroyed
upon scattering with
particle 1. It is
due to a non trivial
cancelation of
terms if 
 momenta $q^+$,
$q^-$ satisfy
the pole condition
for $S^{uu}_{23}$,
that
$S^{ub}_{1(23)}$
indeed leaves $V_A$
invariant and acts as
a pure spin
independent phase
on the wavefunction.
We use the
identities above to
simplify expression
(\ref{equa1}) and
find
\be
S^{ub}_{1(23)}=
\frac{1}{4}[(1+s_-)(1+s_+)+2s_+(1-s_-)]
+P_{13}[1+s_- -3s_+
+ s_+s_-].
\ee
Explicit calculation
shows that the term
multiplying $P_{13}$
vanishes for
$q^+,q^-$ satisfying
the pole condition
(\ref{xik}). It
follows that
$S^{ub}_{1(23)}$ is
indeed proportional
to the identity on
$V_A$, leaving the
bound state
invariant up to a
phase:
\be
S^{ub}_{1(23)}=
\frac{1}{2}(1+s_-)=\frac{\sin
k-\p(q)-i\uv}
{\sin k -\p(q)+i\uv}
\ee
In an analogous
manner one writes
for the S-matrix of
two bound states,
consisting of
particles 1 and 2
parameterized
by $\p(q_{12})$ and 3
and 4, parameterized
by $\p(q_{34})$,
\be
S^{bb}_{(12)(34)}=S^{ub}_{1(34)}S^{ub}_{2(34)}
\ee
and one finds
immediately
\be
S^{bb}_{(12)(34)}=\frac{\p_{12}-\p_{34}-i\uz}
{\p_{12}-\p_{34}+i\uz}.
\ee


\begin{references}
\bibitem{kirkir} The argument actually requires  sharpening: a $Z_2$ symmetry
needs to be respected.
\bibitem{note} It is interesting to note the simplicity of the proof
as compared with the one
in Ref.[1], which extends on 15 pages.
Their main theorem (2.1), stating  that $\eta^+|0\ran$ is not a BA state,
is wrong since the possibility of
complex momenta is not considered.
\end{references}
\end{document}